\documentstyle[11pt]{article}

\newtheorem{theorem}{Theorem}

\newenvironment{proof}%
 {\par\noindent{Proof \quad}}{\hfill$\Box$\bigskip}
\newenvironment{remark}%
 {\par\smallskip\noindent{\underbar{{\it Remark}} \quad}}{\par\smallskip}

\newcommand{\tr}{{\rm Tr}}
\newcommand{\rgl}{\rangle}
\newcommand{\lgl}{\langle}

\begin{document}
\begin{center}
\vspace*{2mm}
{\Large Uhlmann's  parallelism in quantum estimation theory}\\
\vspace{10mm}
Keiji Matsumoto\\
{\it Department of Mathematical Engineering and Information Physics }\\
{\it University of Tokyo, Tokyo 113, Japan }
\end{center}

\begin{abstract}
Two important classes of the quantum statistical model,
the locally quasi-classical model
and the  quasi-classical model, 
 are introduced 
from the estimation theoretical viewpoint,
and they are  characterized geometrically by 
the vanishing conditions of the relative phase factor (RPF),
implying the close tie between
Uhlmann parallel transport and the quantum estimation theory.
\end{abstract}

\section{Introduction, Uhlmann's parallelity, and SLD}
Berry's phase, by far confirmed by several experiments,
is a holonomy of a natural connection on the line bundle
over the space of pure states \cite{AA}\cite{Berry}. 
In 1986, Uhlmann generalized the theory to include
mixed states in the Hilbert space ${\cal H}$
\cite{Uhlmann:1986}\cite{Uhlmann:1992}
\cite{Uhlmann:1993}.
Throughout this paper, for the sake of clarity,
$n\equiv\dim{\cal H}$ is
assumed to be finite,
and the density matrix is strictly positive,
though 
Uhlmann's original theory is free of these assumptions.

Letting $W$ be such a $n$ by $n$ matrix
that $\rho=\pi(W)\equiv WW^{\dagger}$,
$WU$ also satisfies  $\rho=\pi(WU)$ iff $U$ is a unitary matrix.
So, it is natural to see a space 
${\cal W}=\{W|W\in GL(n,{\bf C})\:,{\rm Tr}WW^{\dagger}=1\}$ 
as a fiber bundle over the space of
strictly positive density matrices ${\cal P}_n$ in ${\cal H}$ 
with $U(n)$'s being its fiber.
One possible physical interpretation of
$W$ is a representation of a state vector 
$|\Phi\rgl$ in the bigger
Hilbert space ${\cal H}\otimes{\cal H}'$.
Here, $\dim{\cal H}'$ is $n$ and   
the operation $\pi(*)$ corresponds to  the partial trace of
$|\Phi\rgl\lgl\Phi|$ over ${\cal H}'$.

To introduce a connection \cite{KobayashiN}, 
or a concept of parallel transport 
along the curve $C=\{\rho(t)|t\in {\bf R}\}$ in ${\cal P}_n$,
a  {\it horizontal lift} $\{W(t)|t\in {\bf R}\}\in{\cal W}$ of C
is defined so that  $\rho(t)=\pi(W(t))$ and 
\begin{eqnarray}
\frac{dW(t)}{dt}=\frac{1}{2}L^S_t(t)W(t),
\label{horizontal}
\end{eqnarray}
are satisfied,
where $L^S_t(t)$ is a Hermitian matrix 
is the root of the matrix equation
$d\rho(t)/dt=(1/2)(L^S_t(t)\rho(t)+\rho(t)L^S_t(t))$.

Letting $\{W(t)|t\in{\bf R}\}$ be a horizontal lift of 
$C'=\{\rho(t)|0\leq t \leq 1\}$,
the {\it relative phase factor} (RPF) between  $\rho_0$ and $\rho_1$
along the curve $C$
is the unitary matrix $U$ defined by the equation $W(1)=\hat{W}_1 U$, 
where 
$\hat{W}_1$ 
satisfies $\rho(1)=\pi(\hat{W}_1)$ and 
$\hat{W}_1^{\dagger}W(0)=W^{\dagger}(0)\hat{W}_1$.
RPF is said to vanish when it is equal to the identity.  

Back in the 1968,
Helstrom independently introduced
the Hermitian matrix $L^S_t(t)$,
which played a major role in the definition $(\ref{horizontal})$ 
of Uhlmann's parallelity,
as a key concept of 
his statistical estimation theory of quantum states.
He called the matrix $L^S_t(t)$ {\it symmetrized logarithmic derivative} (SLD)
because  SLD is introduced as  
a quantum counterpart of a logarithmic derivative 
in the classical estimation theory
(throughout this paper, the term `classical estimation' means
estimation of probability distributions)
\cite{Helstrom:1967} \cite{Helstrom:1976}.
Our starting point is the following queries: 
Why SLD  plays such an important role both in the quantum estimation theory
and in Uhlmann's parallelity ? Is this just a coincidence?

\section{Quantum estimation theory}
In this section, conventional theory of quantum estimation is
reviewed briefly.
In the quantum estimation theory,
we try to know the density matrix of the given system 
from the data $\xi\in\Xi$ produced from a measuring apparatus.
For simplicity, it is assumed that the system
belongs to a certain model
${\cal M}
=\{\rho(\theta)|\theta\in \Theta \subset {\bf R}^m\}\subset{\cal P}_n$,
and that  
the true value of the parameter $\theta$ is not known.
For example, ${\cal M}$ is 
a set of spin states with given wave function part and unknown spin part.
An estimate $\hat\theta$ is obtained 
as a function $\hat\theta(\xi)$  of data $\xi\in Xi$ to ${\rm R}^m$.
The purpose of the theory is to 
obtain the best estimate and its accuracy.
The optimization is done by the appropriate choice of
the measuring apparatus and the function $\hat\theta(\xi)$ from
data to the estimate.

Whatever apparatus is used,
the data $\xi\in\Xi$
lie in a particular subset $B$
of $\Xi$ writes
\begin{eqnarray}
{\rm Pr}\{ \xi \in B|\theta \} =\tr \rho( \theta )M(B),
\label{eqn:pdm}
\end{eqnarray}
when the true value of the parameter is $\theta$.
Here, $M$, which is called {\it measurement},
is a mapping  
from subsets $B\subset \Xi$ to non-negative Hermitian matrices in 
${\cal H}$, such that
\begin{eqnarray}
&&M(\phi)=O, M(\Xi)=I,\nonumber\\
&&M(\bigcup_{i=1}^{\infty} B_i),
=\sum_{i=1}^{\infty}M(B_i)
\;\;(B_i\cap B_j=\phi,i\neq j),
\label{eqn:maxiom}
\end{eqnarray}
(see Ref.{\rm \cite{Helstrom:1976}},p.53 
and Ref.{\rm \cite{Holevo:1982}},p.50.).
Conversely, some apparatus corresponds to any measurement $M$
\cite{Stinespring:1955}\cite{Ozawa:1984}.
A pair $(\hat\theta,M,\Xi)$
is called an {\it estimator}. 

An estimator $(\hat\theta,M,\Xi)$ is said to be  {\it locally unbiased}
at $\theta$ 
if 
\begin{eqnarray}
&&E_\theta[\hat\theta(\xi)|M,\Xi]=\theta,\nonumber\\
&&\partial_i E_\theta[\theta^j(\xi)|M,\Xi]=\delta^j_i\:(i,j=1,...,m),
\end{eqnarray}
hold at $\theta$,
where $E_\theta[*|M,\Xi]$ is 
the expectation with respect to the probability measure $(\ref{eqn:pdm})$,
and
$\partial_i$ stands for $\partial/\partial\theta^i$.
Only locally unbiased estimators are treated from now on.

In the classical estimation,
the inverse of so-called Fisher information matrix
provides the tight lower bound of
covariance matrices of locally unbiased estimates,
where local unbiasedness of the estimate is 
defined almost in the same way as in quantum estimation.
Coming back to the quantum estimation, 
\begin{eqnarray}
V_{\theta}[\hat\theta(\xi)|M,\Xi]\geq(J^S(\theta))^{-1}
\label{eqn:mpCR}
\end{eqnarray}
holds true,
{\it i.e.}, $V_{\theta}[\hat\theta(\xi)|M,\Xi]-(J^S(\theta))^{-1}$ 
is non-negative definite
for any unbiased estimator $(\hat\theta,M,\Xi)$
(see (5.4) 
in Ref.{\rm \cite{Holevo:1982}},p.276).
Here,
$V_{\theta}[\hat\theta(\xi)|M,\Xi]$ is 
the covariance matrix 
of $\hat\theta=\hat\theta(\xi)$ with respect to 
the probability measure $(\ref{eqn:pdm})$,
and 
$J^S(\theta)=[J^S_{ij}(\theta)]$, which is
analogically called SLD Fisher information matrix, 
is defined by 
\begin{eqnarray}
J^S_{ij}(\theta)={\rm Re}\tr \rho(\theta)
L_{i} ^S (\theta) L_{j}^S (\theta)\:(i,j=1,...,m),
\end{eqnarray}
where
$L_{i} ^S(\theta)$ is the SLD of parameter $\theta^i$, {\it i.e.},
\begin{eqnarray}
\partial_i \rho(\theta)
=\frac{1}{2}(L_{i} ^S(\theta)\rho(\theta)+\rho(\theta)L_{i} ^S(\theta)).
\end{eqnarray}

The bound $(J^S(\theta))^{-1}$ is one of the bests, 
in the sense that
any Hermitian matrix $A$ such that $A\geq(J^S(\theta))^{-1}$,
is no more  a lower bound.
However, different from the classical case, 
the equality in  $(\ref{eqn:mpCR})$ 
is not attainable except for the case indicated by
the following theorem, which is proved by Nagaoka \cite{Nagaoka:1996}.

\begin{theorem}
The equality in $(\ref{eqn:mpCR})$ is attainable at $\theta$
iff $[L^S_i(\theta),L^S_j(\theta)]=0$ for any $i,j$.
Letting
$|\xi\rgl$ be a simultaneous eigenvector of 
the matrices $\{L^S_j (\theta)|j=1,...,m\}$
and 
$\lambda_i(\xi)$ be the eigenvalue of $L^S_i (\theta)$
corresponding to $|\xi\rgl$,
the equality is attained by the estimator
$(\hat\theta_{(\theta)},M_{(\theta)},\Xi)$
such that
\begin{eqnarray}
&&\Xi=\{\xi| \xi=1,...,n\},\nonumber\\
&&M_{(\theta)}(\xi)=|\xi\rgl\lgl \xi|,\nonumber\\
&&\hat\theta_{(\theta)}^j(\xi)
= \theta^j + \sum_{k=1}^n [(J^S)^{-1}]^{jk} \lambda_k(\xi).
\label{eqn:bests}
\end{eqnarray}
\label{theorem:mpCR}
\end{theorem}
\begin{remark}
In this paper, we focus on  the lower bound $(\ref{eqn:mpCR})$,
and are not concerned with the lower bound of ${\rm Tr}GV[M]$,
which is treated in Refs. \cite{Helstrom:1976}-\cite{Holevo:1982}.
\end{remark}

The model ${\cal M}$ is said to be 
{\it locally quasi-classical} at $\theta$ iff
$L^S_i(\theta)$ and $L^S_j(\theta)$ commute for any $i,j$.
In this case, the bound $(\ref{eqn:mpCR})$ becomes tight
as its classical counterpart is and the analogy of classical estimation
seemingly works well. 
However, this analogy fails in that
the measurement $M_{(\theta)}$ in $(\ref{eqn:bests})$ 
is  dependent on the true value of the parameter,
which is unknown before the estimation.
Hence, we need to adopt the measurement through the process of estimation
using the knowledge about the parameter obtained so far \cite{Nagaoka:1996}. 

Let us move to the easier case,
in which
$L^S_i(\theta)$ and $L^S_j(\theta')$
commute for any $\theta\neq\theta'$,
in addition to being locally quasi-classical at any $\theta\in\Theta$.
Here,
the measurement $M_{(\theta)}$ in $(\ref{eqn:bests})$,
denoted by $M_{best}$ hereafter,
is independent of theta
and is uniformly optimal for all $\theta$
(so is the corresponding apparatus).
We say such a model is {\it quasi-classical} \cite{Yung},
because 
given the optimal apparatus,
the quantum estimation  reduces to
the classical estimation.

\section{Vanishing conditions for RPF}
So far, we have reviewed the conventional theory of 
quantum estimation and Uhlmann's parallelity.
In this section, 
we derive conditions for RPF to vanish, which is used to
characterize 
the classes of model defined in the previous section.
For notational simplicity, the argument $\theta$ is omitted, 
as long as the omission is not misleading.

The RPF for the infinitesimal loop
\begin{eqnarray}
&&\theta=(\theta^1,\theta^2,...,\theta^m)
\rightarrow(\theta^1+d\theta^1,\theta^2,...,\theta^m)
\rightarrow(\theta^1+d\theta^1,\theta^2+d\theta^2,...,\theta^m)
\nonumber\\
&\rightarrow&(\theta^1,\theta^2+d\theta^2,...,\theta^m)
\rightarrow(\theta^1,\theta^2,...,\theta^m)=\theta,
\nonumber
\end{eqnarray}
is calculated up to the second order of $d\theta$
by expanding the solution of 
the equation $(\ref{horizontal})$ 
to that order:
\begin{eqnarray}
I+\frac{1}{2}W^{-1}F_{12}W\;d\theta^1 d\theta^2
+o(d\theta)^2,\nonumber\\
F_{ij}
=(\partial_i L^S_j-\partial_j L^S_i)
-\frac{1}{2}[L^S_i,L^S_j].
\label{eqn:Fij}
\end{eqnarray}
Note that $F_{ij}$ is a `representation' of  the curvature form,
and that RPF  for any closed loop vanishes  iff  $F_{ij}$ is zero
at any point in ${\cal M}$.

\begin{theorem}
RPF  for any closed loop vanishes
iff $[L^S_i(\theta),L^S_j(\theta)]=0$ for any $\theta \in \Theta$.
In other words,
\begin{eqnarray}
F_{ij}(\theta)=0\Longleftrightarrow[L^S_i(\theta),L^S_j(\theta)]=0.
\label{eqn:f0lsls0}
\end{eqnarray}
\label{theorem:f0lsls0}
\end{theorem}

\begin{proof}
If $F_{ij}$ equals zero,  
then both of the two terms in the left-hand side of $(\ref{eqn:Fij})$ 
must vanish,
because the first term is Hermitian and the second term is skew-Hermitian.
Hence, if $F_{ij}=0$, $[L^S_i,L^S_j]$ vanishes.

On the other hand, the identity
$\partial_i\partial_j\rho-\partial_j\partial_i\rho=0$, 
or its equivalence
\begin{eqnarray}
(\partial_iL^S_j-\partial_jL^S_i -\frac{1}{2}[L^S_i,L^S_j])\rho+
\rho(\partial_iL^S_j-\partial_jL^S_i +\frac{1}{2}[L^S_i,L^S_j])=0,\nonumber
\label{eqn:didj}
\end{eqnarray}
implies
that
$\partial_i L^S_j-\partial_j L^S_i$ vanishes if $[L^S_i,L^S_j]=0$,
because 
$\partial_i L^S_j-\partial_j L^S_i$ is Hermitian
and $\rho$ is positive definite.
Thus we see $F_{ij}=0$  if $L^S_i$ and $L^S_j$ commute.
\end{proof}

A model ${\cal M}$ is said to be {\it parallel} 
when the RPF between any two points along any curve vanishes.
From the definition, if ${\cal M}$ is parallel, RPF along
any closed loop vanishes, but the reverse is not necessarily true.
The following theorem is a generalization of Uhlmann's 
theory of $\Omega$-horizontal real plane \cite{Uhlmann:1993}.

\begin{theorem}
The following three conditions are equivalent.
\begin{itemize}
\item[{\rm (1)}] ${\cal M}$ is parallel.
\item[{\rm (2)}] Any element $\rho(\theta)$ of ${\cal M}$ writes
	\begin{eqnarray}
	    \rho(\theta)=M(\theta)\rho_0 M(\theta),
	\end{eqnarray}
     where $M(\theta)$ is Hermitian,  and 
    $M(\theta_0)$ and $M(\theta_1)$ commute 
	for any  $\theta_0, \theta_1 \in\Theta$.
\item[{\rm (3)}]$
		\forall i,j,\:\forall \theta_0,\theta_1\in\Theta,\:\;
		[L^S_i(\theta_0),L^S_j(\theta_1)]=0.
		$
\end{itemize}
\label{theorem:m-inphase}
\end{theorem}
\begin{proof}
Let $W(\theta_t)=M(\theta_t)W_0$ be a horizontal lift of
$\{\rho(\theta_t),t\in{\bf R}\}\subset{\cal M}$.
Then, $W^{\dagger}_0W(\theta_t)=W^{\dagger}(\theta_t)W_0$ implies
$M(\theta_t)=M^{\dagger}(\theta_t)$, and
$W^{\dagger}(\theta_{t_0})W(\theta_{t_1})
=W^{\dagger}(\theta_{t_1})W(\theta_{t_0})$ implies 
$M(\theta_{t_0})M(\theta_{t_1})=M(\theta_{t_1})M(\theta_{t_0})$.
Thus we get $(1)\Rightarrow(2)$. Obviously, the reverse also holds true.
For the proof of $(2)\Leftrightarrow(3)$,
see {\rm Ref. \cite{Yung}}, pp.31-33.
\end{proof}

\section{Uhlmann's parallelity  in quantum estimation theory}
In this section, geometrical structure of ${\cal W}$ is
related to the quantum estimation theory.
First, we imply the statistical significance of natural metric
$\tr \dot{W}\dot{W}^{\dagger}$
in the space  ${\cal W}$.
When $dim{\cal M}=1$, 
the equality in $(\ref{eqn:mpCR})$ is always attainable
(see Refs. \cite{Helstrom:1967}-\cite{Holevo:1982}).
By  virtue of the geometrical identity
\begin{eqnarray}
J^S_t(t)=
	\min_{W(t)\in\pi^{-1}(\rho(t))}
		4\tr
			\frac{dW(t)}{dt} 
			\frac{dW^{\dagger}(t)}{dt}
\label{eqn:mindist}
\end{eqnarray}
(see Refs. \cite{Uhlmann:1992}-\cite{Uhlmann:1993}),
the inequality $(\ref{eqn:mpCR})$ in the case of $dim{\cal M}=1$,
allows natural geometrical interpretation:
the closer 
	two fibers
			$\pi^{-1}(\rho(t))$ and $\pi^{-1}(\rho(t+dt))$
				are,
the harder
	it is
	  to distinguish $\rho(t)$ 
		 from $\rho(t+dt)$.

To conclude the paper, we present the theorems which 
geometrically characterize 
the locally quasi-classical model and
quasi-classical model, described statistically so far,
by the  vanishing conditions of RPF, 
implying the close tie between
Uhlmann parallel transport and the quantum estimation theory.
They are straightforward consequences of
the definitions of the terminologies  and
theorems \ref{theorem:mpCR} -\ref{theorem:m-inphase}.
\begin{theorem}
${\cal M}$ is locally quasi-classical at $\theta$
iff $F_{ij}(\theta)=0$ for any $i,j$.
${\cal M}$ is locally quasi-classical at any $\theta\in\Theta$
iff the RPF for any loop vanishes.
\end{theorem}
\begin{theorem}
 ${\cal M}$ is quasi-classical
iff ${\cal M}$ is parallel.
\end{theorem}

\section*{Acknowledgements}
I would like to thank Dr. Akio Fujiwara, for introducing me
quantum estimation theory, and for helpful discussions.
I am also indebted to Dr. Hiroshi Nagaoka for inspiring discussions
and for checking this manuscript carefully.

\end{document}